\documentclass[12pt]{article}

\usepackage{graphicx}

\def \gsim{\mathrel{\vcenter
     {\hbox{$>$}\nointerlineskip\hbox{$\sim$}}}}

\newcommand{\beq}{\begin{equation}}
\newcommand{\eeq}{\end{equation}}
\newcommand{\beqa}{\begin{eqnarray}}
\newcommand{\eeqa}{\end{eqnarray}}
\newcommand{\beqar}{\begin{eqnarray*}}
\newcommand{\eeqar}{\end{eqnarray*}}

\begin{document}

\thispagestyle{empty}

\hfill{\sc CAFPE-31/04}

\vspace*{-2mm}
\hfill{\sc UG-FT-161/04}

\vspace{32pt}

\begin{center}

\textbf{\Large Cosmogenic neutrinos and signals of TeV gravity in
            \\  air showers and neutrino telescopes}

\vspace{40pt}

J.I.~Illana, M.~Masip, and D.~Meloni

\vspace{12pt}

\textit{Centro Andaluz de F\'\i sica de Part\'\i culas Elementales
(CAFPE)\\ and\\ Departamento de F{\'\i}sica Te\'orica y del
Cosmos}\\ \textit{Universidad de Granada, E-18071, Granada, Spain}\\
\vspace{16pt}
\texttt{jillana@ugr.es, masip@ugr.es, meloni@ugr.es}
\end{center}

\vspace{40pt}

\begin{abstract}

The existence of extra dimensions allows the possibility that the
fundamental scale of gravity is at the TeV. If that is the case,
gravity could dominate the interactions of ultra-high
energy cosmic rays. In particular, the production of microscopic
black holes by cosmogenic neutrinos has been estimated in a number
of papers. We consider here gravity-mediated interactions
at larger distances, where they can be calculated in the eikonal
approximation. We show that for the expected flux of cosmogenic
neutrinos these elastic processes give a stronger
signal than black hole production in neutrino telescopes.
Taking the bounds on the higher dimensional Planck mass $M_{D}$
$(D=4+n)$
from current air shower experiments, for $n=2\ (6)$
elastic collisions could produce up to 118 (34) events per
year at IceCube.
On the other hand, the absence of any signal
would imply a bound of $M_D\gsim 5$ TeV.

\end{abstract}


\newpage

{\bf Introduction.}
We observe extensive air showers produced
when a cosmic ray from outer space hits a nucleon in the upper
atmosphere. The observed events \cite{Anchordoqui:2002hs}
have energies of up to $10^{11}$~GeV,
and their profile and distribution are consistent with a primary
proton of extragalactic origin. In their way to the Earth these
protons would interact
with the CMB photons and produce pions:
\beq
p + \gamma_{2.7{\rm K}} \to \Delta^+\to n+ \pi^+\; (p+\pi^0)\;.
\label{eq1}
\eeq
The flux of cosmogenic neutrinos is created in the decay of the
charged pions, and it will appear correlated with observable
fluxes of nucleons and photons.

Cosmogenic neutrinos are of great interest as probes of new
TeV physics because of two generic reasons. First, they provide large
center of mass energies. Second,
the relative effect of new physics on the weakly interacting neutrinos
is larger than on quarks or charged leptons,
making it easier to see deviations.
The signals of new physics could be detected in deeply
penetrating air showers and neutrino telescopes.

In particular,
in models with extra dimensions and the fundamental Planck scale
at the TeV \cite{ADD} the gravitational interactions are unsuppressed
in the transplanckian regime. The possibility of black hole (BH)
formation \cite{BH0} by cosmogenic neutrinos has been discussed in several
papers \cite{feng}-\cite{olinto}.
Here we will study the gravitational interaction
at larger distances, where it can be calculated
using the eikonal approximation \cite{eik,emparan,riccardo}.
This approximation involves linearized
gravity and is not affected by the uncertainties in the cross section
for BH formation. After discussing the bounds on the gravitational
scale from air shower experiments,
we will show that these elastic processes
are more frequent than BH formation in neutrino telescopes.
At large impact parameters the neutrino interacts, loses a small
fraction of energy and keeps going. Telescopes could
detect these processes because they are sensitive
to events of energy three or four orders of magnitude below
the typical energy (around $10^{9}$~GeV) of the cosmogenic neutrinos.

{\bf TeV gravity.}
The simplest picture of TeV gravity includes only two free parameters:
the value of the higher dimensional Planck scale $M_D$, and the number
$n$ of compact dimensions
where gravity propagates. A third parameter, the
(common) length $2\pi R$ of the $n$ dimensions, could be deduced from the
4d Newton constant:
\beq
G_D= (2\pi R)^n G_N = {(2\pi)^{n-1}\over 4 M_D^{n+2}}\;.
\label{eq2}
\eeq
At processes below $M_D$ the model-independent signature of
extra dimensions is graviton emission.
The amount of energy radiated
would be proportional to the accessible phase space or, in the
Kaluza-Klein (KK) picture, to the number of KK modes of mass
below the center of mass energy. In this type of experiments
for a given $n$ one sets bounds on $R$ and then deduces the
limits on $M_D$.
From collider experiments one obtains $M_D\ge 1.4\;(1.0)$ TeV
for $n=2\;(\ge 3)$ \cite{colliders}, whereas from
SN1987A the bounds go up to 22 TeV for $n=2$ \cite{SN}.
One should keep in mind, however, that the gravitons emitted in
the supernova explosion have a KK mass below $\approx 50$ MeV. The simple
picture could be modified above this energy, for example, with
four more dimensions at $R'\sim (100$~GeV)$^{-1}$, which would bring
the fundamental scale of gravity down to 1 TeV without affecting
the physics in the supernova.

The bounds obtained from transplanckian collisions are
complementary in the sense that given $n$ they are a direct probe of
$M_D$, and $R$ is then adjusted in order to reproduce $G_N$.
At energies above $M_D$ and impact parameters smaller than $R$
the collision is a pure higher-dimensional process independent
of the compactification details that fix the value of
the effective Newton constant. The transplanckian
collision does not {\it see} that the extra dimensions
are compact, they could be taken infinite
with no effect on the cross section.

{\bf Neutrino-nucleon cross section.}
The TeV gravity model should be embedded in a string theory, which
would relate $M_D$ with the string scale $M_S$. In the simplest
set-up \cite{peskin} the standard
model (SM) fields (open strings) would be attached to
a 4d brane, whereas gravity (closed strings) would propagate
in the whole $D$d space. In this case
\beq
M_D^{n+2}={8 \pi\over g^4} M_S^{n+2}\;,
\label{eq3}
\eeq
with $g$ the string coupling. The transplanckian regime
corresponds then to energies above the string scale, where
any tree-level amplitude becomes very {\it soft}. In the
ultraviolet string amplitudes go to zero exponentially
at fixed angle and, basically, only the forward (long distance)
contribution of the graviton survives.
This is precisely the regime
where the eikonal approximation is valid.

Let us consider the elastic collision of a neutrino and a
parton that exchange $D$-dimensional gravitons (see
\cite{emparan,riccardo} for details).
The eikonal amplitude ${\cal A}_{eik}(s,t)$ resums the infinite
set of ladder and cross-ladder diagrams.
It is reliable as far as
the momentum carried by the gravitons is smaller than
the center of mass energy or, in terms of the fraction
of energy $y=(E_\nu-E'_\nu)/E_\nu$ lost by the incoming neutrino,
if $y=-t/s\ll 1$ ($s$ and $t$ refer to the Mandelstam
parameters at the parton level). In this limit the
amplitude is independent of the spin of the colliding particles.
Essentially, ${\cal A}_{eik}$ is the exponentiation of the Born
amplitude in impact parameter space:
\beq
{\cal A}_{eik}(s,t)={2 s\over i}
\int {\rm d}^2b\; e^{i\mathbf{q}\cdot\mathbf{b}}\;
\left(e^{i\chi (s,b)}-1\right)\;,
\label{eq4}
\eeq
where $\chi (s,b)$ is the eikonal phase
and $\mathbf{b}$ spans the (bidimensional) impact parameter space.
$\chi (s,b)$ can be deduced from the Fourier transform to
impact parameter space of ${\cal A}_{Born}(s,t)$. 
Our Born amplitude comes from the $t$-channel exchange
of a higher dimensional graviton:
\beq
{\cal A}_{Born}=-{s^2 \over M_D^{n+2}}
\int {{\rm d}^n\;q_T\over t-q_T^2} \;,
\label{eq5}
\eeq
where the integral over momentum $q_T$ along the extra dimensions
(equivalent to the sum over KK modes) gives an UV
divergence. The {\it magic} of the eikonal amplitude is
that it will be well defined despite we obtain it from an UV
dependent Born amplitude: the contributions from large $q_T$ introduce
corrections to the phase $\chi (s,b)$ only at small $b$
($\approx 1/q_T$), but this small $b$ region gives
a negligible contribution to ${\cal A}_{eik}$ in the transplanckian
regime.

From ${\cal A}_{eik}$ one obtains
$\chi(s,b)=(b_c/ b)^n$, with
\beq
b_c^n=
{(4\pi)^{{n\over 2}-1}\over 2}\Gamma\left({n\over 2}\right)
{s\over M_D^{2+n}}\;.
\label{eq6}
\eeq
The amplitude in Eq.~(\ref{eq4}) can \cite{emparan,riccardo}
then be written as
${\cal A}_{eik}(s,q)=4\pi s b_c^2\; F_n(b_c q)\;$, where
\beq
F_n(y)=-i
\int_0^\infty {\rm d}x\;x\; J_0(xy)
\left( e^{ix^{-n}} -1 \right)\;,
\label{eq7}
\eeq
$q=\sqrt{-t}$, and the integration variable is $x=b/b_c$.
For $q<b_c^{-1}$ this integral is dominated by impact parameters
around $b_c$, and for $q>b_c^{-1}$ by
a saddle point at $b_s$.
As $q$ (or $y=q^2/s$) grows nonlinear corrections
(H diagrams) become important \cite{riccardo}. For $-t/s\approx 1$
$b_s$ approaches \cite{emparan} the
Schwarzschild radius $R_S$ of the system:
\beq
R_S=\left({2^n\pi^{n-3\over 2}\Gamma\left({n+3\over 2}\right)
\over n+2}\right)^{1\over n+1}
\left({s\over M_D^{2n+4}}\right)^{1\over 2(n+1)}\;.
\label{eq8}
\eeq
At $b\le R_S$ one expects an inelastic collision,
with a significant emission of gravitons, and
black hole (BH) formation. The latter possibility has been
considered in several analyses \cite{BH0}-\cite{olinto},
where is it also shown that a number of factors
(angular momentum, charge,
geometry of the trapped surface, radiation before the collapse)
make a quantitative estimate difficult.
In particular, the higher curvature corrections discussed in
\cite{rychkov} could affect the evolution of the collision after the
horizon has formed, making the simple picture of single BH
production and subsequent Hawking evaporation unlikely.

The differential cross section ${\rm d}\sigma_{eik}/{\rm d} y$
grows as $y$ decreases \cite{emparan}. For example, taking
$M_D=1$ TeV and $E_\nu=10^{10}$~GeV, for $n=2\;(6)$ it is a
factor of 265 (62) larger at $y=10^{-3}$ than at
$y=0.1$. The small $y$ region corresponds to long distance processes
where the neutrino interacts with a parton and transfers only a small
fraction of its energy.
This region is less important for a larger number of
extra dimensions, since then gravity {\it dilutes} faster and becomes
weaker at long distances.
On the other hand, values of $y$ close to 1 mean shorter distance
interactions. For example, we obtain that for $y=0.5$ a $52\%$ of
the $\nu N$ eikonal cross section comes from impact parameters
$b<R_S$ for $n=2$ (or a $71\%$ for $n=6$).
We will then separate two types of transplanckian
($s>M_D^2$ at the parton level) processes:

{\it (i)} Inelastic processes where
the neutrino interacts with a parton at distances
$b\le R_S$. The cross section for these processes,
$\sigma_{BH}=\pi R_S^2$, would
include BH formation and {\it hard} scatterings with
important graviton emission where 
the neutrino loses most of its initial energy.

{\it (ii)} Elastic processes where the
neutrino transfers to the parton a small fraction $y$ of its
energy (we take $y_{max}=0.2$), and keeps going. We use the
eikonal approximation to describe these processes. 
They are dominated by impact parameter distances 
larger than $R_S$ and thus non-linear effects and graviton
emission are expected to be small.

{\bf Air showers from cosmogenic neutrinos.}
The flux of cosmogenic neutrinos depends on the production
rate of primary nucleons. It will appear correlated with proton
and photon fluxes that should be consistent, respectively,
with the number of ultrahigh energy events at AGASA and
HiRes \cite{Anchordoqui:2002hs} and with the
diffuse $\gamma$-ray background measured by EGRET
\cite{Sreekumar:1997un}.
We will consider two neutrino fluxes described in \cite{semikoz}.
The first one saturates the observations by EGRET, whereas for
the second one the correlated flux of $\gamma$-rays contribute
only a 20\% to the data, with the nucleon flux normalized
in both cases to AGASA/HiRes. The higher
flux predicts 820 down-going neutrinos of each flavor with energy between
$10^8$~GeV and $10^{11}$~GeV per year and km$^2$, versus
370 for the lower one.

AGASA and Fly's Eye are able to detect efficiently 
penetrating air showers
of energies above $\approx 10^{10}$~GeV \cite{feng}.
When a cosmogenic neutrino of energy
$E_\nu \gsim 10^{10}\;{\rm GeV}$
enters the atmosphere it can experience the two types of processes
described above. Short distance collisions
may produce a BH, whose thermal evaporation would start an air
shower of energy (for negligible graviton emission) up to $0.8 E_\nu$.
On the other hand, if the neutrino suffers
a long distance collision it will transfer to the
parton a fraction $y$ of energy 
up to $y_{max}=0.2$. These processes, well described by the
eikonal approximation, start hadronic showers of energy $y E_\nu$.

We have calculated the combined number of events at AGASA
and Fly's Eye as a function of $M_D$ (we take the exposures to
penetrating showers from \cite{feng}).
Within the SM one expects 0.03 (0.009 for the lower flux)
deeply penetrating showers started by a neutrino,
with a background of 1.72 events from
hadronic showers. In these experiments 1 event passes all the cuts,
which implies \cite{feng} an upper bound of 3.5 neutrino
events at $95\%$ CL. For the higher flux and $n=2$ we obtain 3.5
events (2.1 BH and 1.4 elastic) if $M_D=1.0$ TeV, whereas for
$n=6$ we have 2.6 BH plus 0.9 eikonal events if $M_D=1.5$ TeV.
Neglecting the short distance collisions, for $n=2$ (6)
we obtain 3.5 elastic events for $M_D=0.7\; (0.9)$ TeV.
For the lower flux, the 3.5 events limit would be obtained
(from elastic processes only) for $M_D=0.4$ TeV if $n=2$ or
for $M_D=0.5$ TeV if $n=6$.

{\bf TeV gravity events at IceCube.}
IceCube \cite{icecube} is a large scale (km$^3$) neutrino telescope
currently under construction in the Antarctic ice. Its center
is at a depth of 1.8 km, which implies that if
$\sigma_{\nu N}\le 0.01$ mb neutrinos can reach it vertically
with no previous interactions, whereas if
$\sigma_{\nu N}\le 0.0001$ mb they could also reach it horizontally
after crossing 150 km of ice. The detector is sensitive to
hadronic showers of energy $E_{sh} > 500$ TeV.

To be definite, let us consider a cosmogenic neutrino
of energy $E_\nu=10^{10}$~GeV for $M_D=1$~TeV and $n=2\;(6)$.
The probability that the neutrino survives to reach the detector
from a zenith angle $\theta_z$ is
\beq
P_{surv}=\exp[-X(\theta_z)\;\sigma\;N_A]\;,
\label{eq9}
\eeq
where $X(\theta_z)\approx \rho_{ice} L(\theta_z)$ is the column
density of material ($L(\theta_z)$ is
the length of the column in ice) in its way to the detector,
and $\sigma=\sigma_{BH}+\sigma_{SM}$ is the inelastic
cross section (we do not include $\sigma_{eik}$ because
the elastic processes with $y < 0.2$ introduce a negligible
distortion in the energy of the neutrinos that reach the
detector).
For $\cos \theta_z\approx 0.11\ (0.45)$
the length $L(\theta_z)$
is equal to its mean free path $L=1/(\rho_{ice}\sigma N_A)$,
therefore the neutrino should typically reach the detector
from smaller angles. Within the SM this angle would
go up to $\cos \theta_z\approx -0.03$. Once in the detector,
the probability that the neutrino experiences a short distance
interaction is given by
\beq
P_{int}^{BH}=1-\exp[-L\;\rho_{ice}\;\sigma_{BH}\;N_A]\;,
\label{eq10}
\eeq
where $L\approx 1$ km is the linear dimension of the detector.
We obtain values of $P_{int}^{BH}$ from 0.06 for $n=2$ to 0.2 for
$n=6$.
On the other hand, the probability $P_{int}^{eik}$ that the
neutrino interacts elastically, loses a fraction of energy
between $y_{min}=(500\;{\rm TeV})/E_\nu=5\times 10^{-5}$
and $y_{max}=0.2$, and starts an observable hadronic shower,
can be read from the expression above just by changing
$\sigma_{BH}$ by ${\rm d}\sigma_{eik}/{\rm d}y$ integrated
between $5\times 10^{-5}$ and $0.2$.
We obtain that $P_{int}^{eik}$ goes from 0.4 for $n=2$ to 0.3
for $n=6$. These probabilities only change a 1\% 
if we take $y_{max}/2$ or $2 y_{max}$. Within
the SM the probability that the neutrino starts a
shower inside the detector is just $P_{int}^{SM}=0.002$.

Given a cosmogenic neutrino flux $\Phi_{\nu}$, the
number $N_{\rm sh}$ of shower events at IceCube can be estimated as
\beq
N_{\rm sh}=\sum_{i} 2\pi A T \int {\rm d} \cos\theta_z
\int {\rm d}E_{\nu} {{\rm d}\Phi_{\nu}\over dE_\nu} P_{surv} P_{int}\;,
\label{eq12}
\eeq
where the sum goes over the three neutrino and antineutrino
species, $A\approx 1$ km$^2$ is the detector's
cross sectional area with respect to the $\nu$ flux, and $T$
is the observation time. For the higher (lower) flux,
within the SM we obtain 1.4 (0.5) hadronic
or electromagnetic events (muons and taus do not shower)
per year above $500$ TeV.

We now consider values of $M_D$ above the bounds
obtained from the absence of penetrating air showers
and calculate the number of events per year at IceCube.
Taking the higher flux, for $n=2\;(6)$ we obtain a maximum of
118 (34) elastic events versus just 20 (24) short distance
events. The elastic events correspond to soft processes
with $y<0.2$. In particular, for $n=2$, 95 of the 118
showers have an energy below $10^{8}$ GeV, versus just
0.08 of the 20 showers from short distance interactions.
For $n=6$, 18 of the 34 eikonal events are
below $10^{8}$ GeV.

\begin{figure}
\centerline{\includegraphics[width=0.7\linewidth]{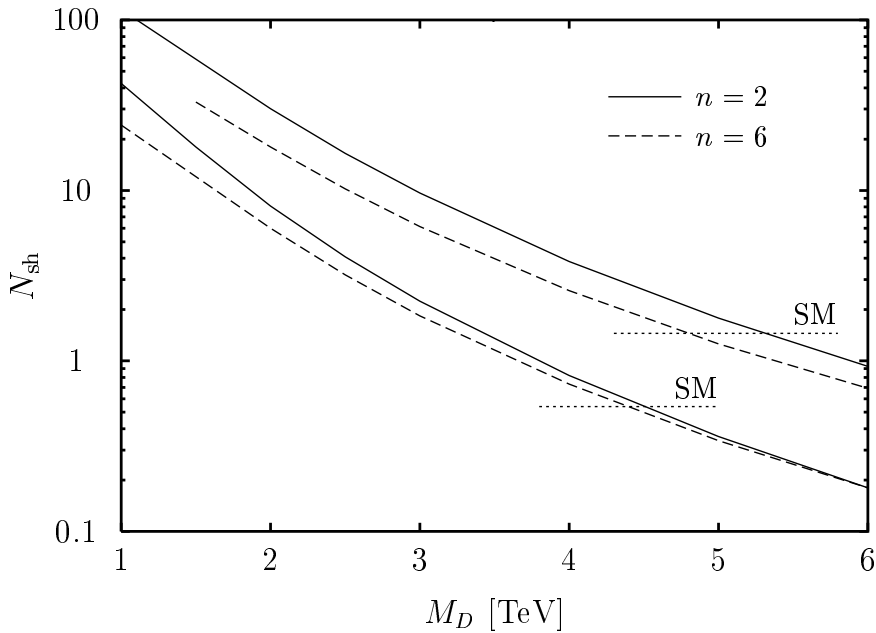}}
\caption{
Number of (eikonal and SM)
shower events per year at IceCube for the two
cosmogenic neutrino fluxes.
\label{fig1}}
\end{figure}

In Fig.~\ref{fig1} we plot the number of elastic events per year
as a function of $M_D$ for the two cosmogenic fluxes.
We find that IceCube could detect
TeV gravity effects above the SM background
for $M_D$ up to approximately 5~TeV. The characteristic signature
would be always a hadronic shower, as charged leptons are never
produced in the gravitational interaction starting the shower.

{\bf Discussion.}
Cosmogenic neutrinos interact with the terrestrial nucleons
at center of mass energies
$\sqrt{2m_N E_\nu}\approx 10^5$~GeV, so they could be used as
probes of new TeV physics. In particular, the possibility
of BH formation in models with extra dimensions has been
entertained by several groups.
These analyses are based on a geometric cross section
that assumes single BH production whenever the neutrino and the
parton interact at impact parameters smaller than $R_S$.

The problem with this estimate is that, despite the large
energy of cosmogenic neutrinos, the $\nu N$ cross section is
dominated by the small $x$ region, and most of the BHs produced
in penetrating air showers or neutrino telescopes would be very
light, with masses just above $M_D$.
These {\it light} BHs would be very sensitive to effects like
graviton emission during the collapse, higher curvature
corrections, or non-thermal effects in the evaporation,
which add uncertainty to the estimate.

In this paper we have analyzed a different type of signal.
It is produced when the neutrino interacts
elastically with the parton at typical distances larger
than $R_S$ and transfers a small fraction $y$ of its
energy. The process is properly described by the eikonal
approximation, and its distinct
experimental signature would be a hadronic
shower of energy $yE_\nu$. Electromagnetic 
showers would never be produced in the initial $\nu N$
interaction.

We have computed the number of penetrating air showers from soft
($y\le 0.2$) elastic processes and from short distance ($b< R_S$)
processes, and have found that the expected number of events at AGASA
and Fly's Eye from these two types of processes is similar.
We obtain bounds on $M_D$ that are below the ones from
SN1987A for $n=2$, but are similar to the limits from collider
experiments for any value of $n$.

In contrast, we have shown that these elastic
interactions provide a clear
and model-independent signal of TeV gravity that would
dominate over BH production in neutrino telescopes.
The reason is that telescopes are sensitive
to showers of energies up to four orders of magnitude below the
average energy of cosmogenic neutrinos.
The excess of hadronic showers
at IceCube could be observed for $M_D$ up to 5 TeV.

We would like to thank Eduardo Battaner and
Roberto Emparan for valuable conversations.
This work has been supported by MCYT (FPA2003-09298-C02-01) and
Junta de Andaluc\'\i a (FQM-101). J.I.I.~and D.M.~(M.M.)
acknowledge financial support from the
European Community's Human Potential Programme HPRN-CT-2000-00149
(HPRN-CT-2000-00152).

\end{document}